\documentclass[journal=prl,showpacs,showkeys,twocolumn,nofootinbib,notitlepage]{revtex4-2}

\pdfpagewidth=\paperwidth
\pdfpageheight=\paperheight

\usepackage{amsmath,graphicx}
\usepackage{verbatim}
\usepackage{color}
\usepackage{amsfonts,amsbsy,amssymb}

\def\L{\mathcal L}

\def\brho{\boldsymbol \rho}

\def\f{\boldsymbol {\rm f}}
\def\l2{\ell_2}

\def\dx{{\rm d}x\,}
\def\dy{{\rm d}y\,}
\def\dz{{\rm d}z\,}
\def\k{\boldsymbol {\rm k}}
\def\f{\hat {\boldsymbol {\rm f}}}
\def\X{\boldsymbol {\rm X}}
\def\T{\boldsymbol {\rm T}}
\def\sinc{{\rm sinc}}
\def\S{\boldsymbol {\rm S}}
\def\d{\boldsymbol {\rm d}}
\def\lam{\boldsymbol \Lambda}
\def\ff{ {\mathfrak a}}
\def\FF{{\mathcal F}}

\def\Th{\boldsymbol \Theta}

\begin{document}

\author{Paul N. Patrone}
\email{paul.patrone@nist.gov}
\author{Matthew DiSalvo}
\author{Anthony J. Kearsley}
\author{Geoffrey B. McFadden}
\author{Gregory A. Cooksey}
\affiliation{National Institute of Standards and Technology \\ 100 Bureau Drive, Gaithersburg, MD 20899, USA}

\date{\today}
\title{Reproducibility in Cytometry: \\ Signals Analysis and its Connection to Uncertainty Quantification}

\begin{abstract}
Signals analysis for cytometry remains a challenging task that has a significant impact on uncertainty.  Conventional cytometers assume that individual measurements are well characterized by simple properties such as the signal area, width, and height.  However, these approaches have difficulty distinguishing inherent biological variability from instrument artifacts and operating conditions.  As a result, it is challenging to quantify uncertainty in the properties of individual cells and perform tasks such as doublet deconvolution.  We address these problems via signals analysis techniques that use scale transformations to: (I) separate variation in biomarker expression from effects due to flow conditions and particle size; (II) quantify reproducibility associated with a given laser interrogation region; (III) estimate uncertainty in measurement values on a per-event basis; and (IV) extract the singlets that make up a multiplet.  The key idea behind this approach is to model how variable operating conditions deform the signal shape and then use constrained optimization to ``undo'' these deformations for measured signals; residuals to this process characterize reproducibility.  Using a recently developed microfluidic cytometer, we demonstrate that these techniques can account for instrument and measurand induced variability with a residual uncertainty of less than 2.5\% in the signal shape and less than 1\% in integrated area.  
\end{abstract}

\maketitle

\section{Introduction}

In the last 30 years, cytometry has evolved as a powerful technique for clinical diagnostics, drug development, and biotechnology \cite{Review1,Review2}.  This success has recently motivated fundamental questions and studies designed to better understand the  ultimate capabilities of cytometers, as well as their potential for quantitative, reproduceable, and even traceable measurements \cite{Cytoref1,Reproducibility1,Reproducibility2,Reproducibility3}.  However, in-depth uncertainty quantification (UQ) has yet to be fully realized, limiting efforts to refine the metrology aspects of cytometry.

In this context, signals analysis is a challenging and often overlooked task that has a significant impact on uncertainty.  For example, conventional flow cytometers assume that individual events are well characterized by simple properties such as the signal area, height, and width \cite{Sigs1,Sigs2,Sigs3,Sigs4}.  However, such approaches discard the vast majority of information in the measurement.  Moreover, cytometers only interrogate a particle once per laser region, so that population variability is inherently convolved with other sources of uncertainty.  As a result, it is difficult to determine whether an event corresponds to a valid measurand, characterize reproducibility on a {\it per-event basis}, and ultimately, quantify confidence in the measurement process.  

The goal of this manuscript is to address such problems by incorporating UQ directly into the signals analysis.  We propose a collection of techniques that use scale transformations to identify sources of variation in events arising from changes in particle speed, size, and brightness.  The main idea behind these analyses is to represent signals in terms of low-order mathematical interpolations and define the relevant transformations in terms of generic physical models.  We then use constrained optimization to map all events onto one another, which also yields the physical parameters (e.g.\ size) associated with the particles.  Using a recently developed microfluidic cytometer (see Ref.\ \cite{Part1}), we illustrate how this analysis can: (i) quantify phenomena such as noise in flow conditions and sample variability; and (ii) estimate the {\it per-event} reproducibility associated with all remaining sources of uncertainty.  We also demonstrate that this analysis provides a foundation for more advanced signal processing techniques by using it to extract the individual singlets comprising a multiplet signal, i.e.\ an event composed of multiple overlapping singlets.  


The need to incorporate UQ directly into signals analysis arises from several issues unique to cytometers.  For example, cells are only  measured once per laser region; moreover, it is difficult to ensure identical flow and optical conditions for all measurands.  Thus, it is impossible to directly characterize repeatability and reproducibility of any measurement.   Signals analysis and mathematical modeling therefore take on new roles, since they allow us to answer the questions, ``how does variation in measurement conditions deform the signals, and how can these deformations be undone?''  In this way, the theory approximately reconstructs an imaginary scenario in which one could repeat measurements on identical particles, thereby overcoming experimental limitations.

Multiplets provide a similar motivation for our analysis techniques.  Because engineering solutions alone cannot prevent multiplets \cite{Doublets1,Doublets2,Doublets3,Doublets4}, common practice discards their information.  However, this introduces uncertainty into population variability estimates \cite{Doublets1,Doublets3}.  Perhaps worse, common data analysis strategies that rely on integrated areas cannot distinguish dim doublets from bright singlets, for example.  This leads to an uncomfortable situation in which even the identity and number of measurands have potentially unquantifiable uncertainties.  Again, we demonstrate that signals analysis is an appropriate avenue to address such problems, since it allows us to extract individual singlets by understanding how they combine to make the multiplet.  
More generally, these examples illustrate that signals analysis is a useful tool for untangling sources of uncertainty that become intertwined by the measurement process.  

This example of {\it multiplet deconvolution} highlights another theme of our work: there is a natural feedback between UQ and signals analysis that must be exploited to realize the full potential of cytometry.  While it is obvious that uncertainty estimates arise from data analysis, we also illustrate the reverse: {\it new data analyses are enabled by UQ.}  For example, we recast multiplet deconvolution as the task of finding the most probable singlets that reconstruct the original multiplet.  Thus, estimating the uncertainty in the shape of singlets is a prerequisite step.   Similar arguments apply to the related but distinct problem of {\it multiplet detection.}  More generally, UQ can inform data analysis by ensuring that results are physically meaningful.\footnote{These observations clarify our stance that UQ is the broad set of tasks that increase confidence in a measurement.}

Because the present work develops new analysis tools, it is useful to perform validation measurements on particles with well characterized properties.  For this reason, the examples below are restricted to fluorescent microspheres.    However, extensions to cell-based measurements are straightforward, if not trivial, and we point to relevant issues throughout.  We note also that the current manuscript does not consider issues associated with gating strategies and quantification of population variability per se.  Such topics are left for future work.  

The reader should also note that while our analysis strives to be as general as possible -- for example, we need not specify a form of the laser profile -- we make certain assumptions that may not apply to all cytometers.  These relate to uniformity of the laser profile perpendicular to the flow direction and light-collection geometric factors.  Importantly, these assumptions allow us to formulate the scale transformations as relatively simple, closed-form expressions, thereby facilitating downstream numerical optimization.  An alternative would be to tailor our approach to a specific cytometer through more detailed modeling, but this may introduce significant computational overhead.  Such issues are discussed in Sec.\ \ref{sec:discussion}.  

The rest of the manuscript is organized as follows.  Section \ref{sec:theory} presents the main assumptions of our analysis in the context of a generic model of a cytometry measurement (\ref{subsec:model}); derives the associated scale transformations (\ref{subsec:transformations}); and formulates the optimization problem needed to undo deformations associated with variable measurement conditions (\ref{subsec:matching}).  Section \ref{sec:validation} validates these methods using experimental data.  Section \ref{sec:applications} highlights the usefulness of this analysis for UQ (\ref{subsec:UQ}) and multiplet deconvolution (\ref{subsec:doublet}).  Section \ref{sec:discussion} considers this work in the greater context of cytometry and signals analysis thereof.  An appendix presents technical aspects of the multiplet deconvolution.

\section{Motivation and Main Ideas}
\label{sec:theory}

A key observation motivates our work: under generic conditions and for a given cytometer, {\it the shapes of all signals are identical up to a set of linear transformations that depend only on the particle properties.}  \textcolor{black}{As will become clear, these transformations can be interpreted as changes of units that bring the numerical values of different measurements into agreement.}  This provides a quantitative framework for comparing events, e.g.\ to determine if they are valid measurands, characterize their properties, and estimate variation about mean behavior.  The goal of this section is to specify the conditions under which this observation holds and develop the tools needed to compare signals.

\textcolor{black}{A key challenge in formulating this analysis is that the measurements are given in a form that does not permit direct comparison.  Part of our task is therefore to convert the signals we can acquire into ones we can use.  The multitude and complexity of steps involved motivates a separation of the analysis into three subsections.  In Sec.\ \ref{subsec:model}, we propose a model that characterizes deformation in signals arising from differences in particle speed and size.  Critically, the particles are assumed to have the same trajectory.  In Sec.\ \ref{subsec:transformations}, we develop the machinery to convert signals acquired over the same time interval (which is experimentally feasible) to ones having the same trajectory length.  Section \ref{subsec:matching} combines these results to solve the backwards problem: determine the particle properties by undoing the signal deformations. }

\subsection{Physical Effects Causing Signal Variation}
\label{subsec:model}

\begin{figure}
\includegraphics[width=8cm]{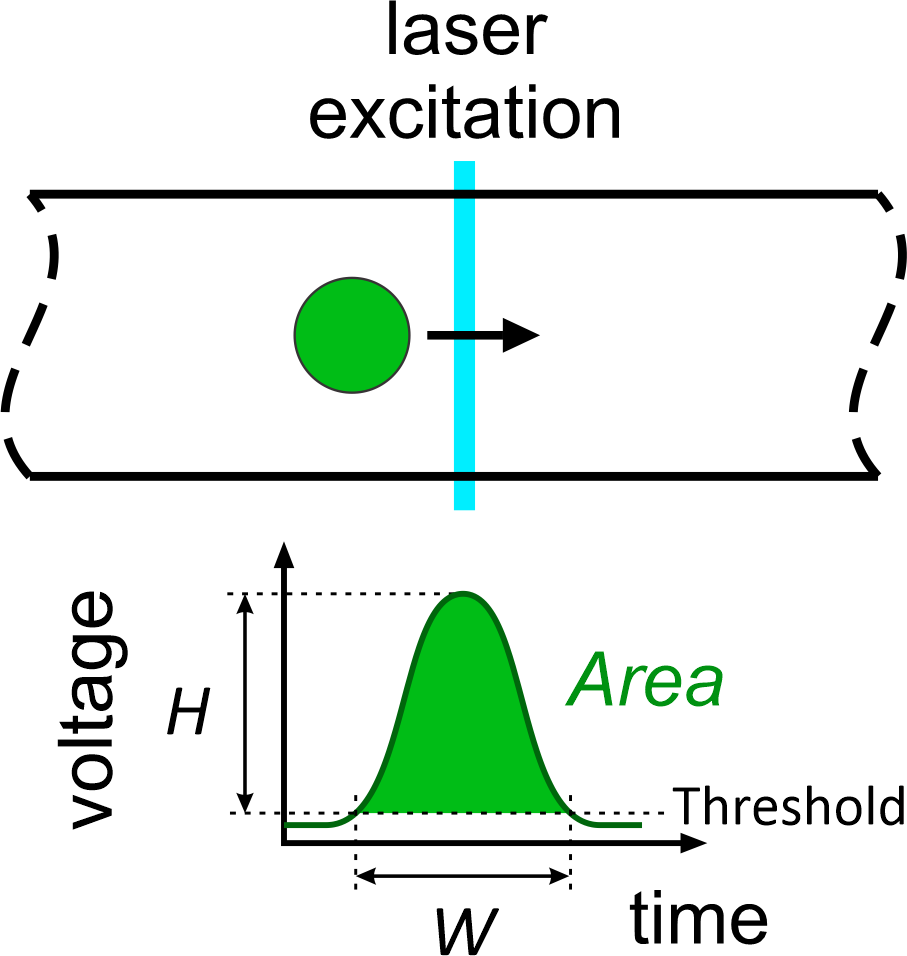}\caption{Schematic of the signal generation process associated with a cytometry event.  As a bead or cell traverses the laser profile, the concentration of fluorophores and object shape convolve with the laser profile to create fluorescent light. This light is converted to a voltage by a photodetector.   See Eq.\ \eqref{eq:genmod} for a mathematical description of this process.  Conceptually we can imagine the signal as beginning and ending when the particle reaches an initial and final position as it moves with the fluid.  We refer to this as a constant-trajectory signal.  In practice, it is easier to acquire a signal over a constant time interval, i.e.\ by approximately centering the peak in a time window of fixed duration, without regard to the distance traveled.}\label{fig:motivational}
\end{figure}

Consider the interplay between physical, chemical, and biological processes during a cytometry measurement.  We take an event to be the fluorescence signal $f(t)$ collected as a function of time $t$ while a bead or cell passes a laser excitation region; see Fig.\ \ref{fig:motivational}.  For particles moving with a constant velocity and parallel to the fluid flow, a model of this process is
\begin{align}
f(t) =\! \int_{D_L} \!\!\!\! \dx\dy\dz C(x,y,z-vt)\Psi(x,y,z)\Phi(x,y,z) \label{eq:genmod}
\end{align}
where $C(x,y,z-vt)$ is the concentration of fluorophores on the particle, $\Psi(x,y,z)$ is the laser light intensity, $\Phi(x,y,z)$ is the amount of \textcolor{black}{emitted} fluorescent light (per unit laser light and fluorophore number) \textcolor{black}{coming from} $(x,y,z)$ and collected by the photodetector,  $v$ is the constant advection velocity of the particle, $D_L$ is \textcolor{black}{the spatial domain over which the signal is generated},  and $z$ is parallel to the direction of flow, i.e.\ the axial direction.  The goal of our analysis is to quantify how $f(t)$ changes as a function of the bead radius $R$, velocity $v$, and concentration $C$, so that we can then solve the reverse problem of determining these parameters from $f(t)$.

We first simplify Eq.\ \eqref{eq:genmod}.  Considering, for example, the microfluidic cytometer in Ref.\ \cite{Part1}, we assume that Eq.\ \eqref{eq:genmod} can be reduced to
\begin{align}
f(t) = \int_{D_L} \dx\dy\dz C(x,y,z-vt)\psi(z) \label{eq:redmod}
\end{align}
where $\psi(z)$ depends only on $z$ and accounts for both the laser and collection optics.  In an abuse of terminology, we henceforth refer to $\psi(z)$ as the laser profile.

Equation \eqref{eq:redmod} is valid for a wide range of operating conditions.  For example, letting $\boldsymbol \rho = (x,y)$ denote the \textcolor{black}{radial} direction, Eq.\ \eqref{eq:redmod} applies when: (i) differently sized, radially symmetric (in $\brho$) particles on a single streamline encounter a laser profile and geometric factors whose product is linear in $\boldsymbol \rho$; and (ii) arbitrary shaped particles on any streamline encounter a laser profile and geometric factors that are independent of $\boldsymbol \rho$.  Either case may happen when the particles are small relative to the laser interrogation region.\footnote{Case (i) arises from symmetry arguments applied to Eq.\ \eqref{eq:genmod}.}   See Ref.\ \cite{Part1} for relevant details associated with the device studied in this manuscript, including in particular the focusing strategy used to ensure that particles do not cross streamlines.

We model $C(x,y,z)$ as a step function of the form
\begin{align}
C(x,y,z) = c \Theta\left(R^2 - x^2 - y^2 - z^2\right) \label{eq:fullbead}
\end{align}
where $c$ is a constant fluorophore concentration and the Heaviside function is defined such that $\Theta(r)=1$ if $r > 0$ and $\Theta(r)=0$ otherwise.  Equation \eqref{eq:fullbead} corresponds to a bead or cell with a uniform concentration of fluorophores throughout its volume.  Through appropriate modification of the Heaviside function this expression can describe, for example, surface concentrations.  Moreover, Eq.\ \eqref{eq:fullbead} can be parameterized to model non-radial deformations associated with deformable cells, although such tasks are beyond the scope of this manuscript.  

In light of these simplifications, Eq.\ \eqref{eq:genmod} becomes
\begin{align}
f(t) &= c\pi\!\! \int_{-L_0}^{L_0} \!\!\!\!\!\dz \!\left[R^2 \!-\! (z\!-\!vt)^2 \right]\!\Theta\!\left[R^2 \!-\! (z\!-\!vt)^2\right]\!\psi(z) \label{eq:modelreduction}
\end{align}
where where we assume that the laser is fully contained inside the domain $[-L_0,L_0]$ for some length $L_0$.  For later convenience, we assume (without loss of generality) that $L_0$ satisfies the inequality
\begin{align}
 0 < a < L_0-R, \label{eq:inequality}
\end{align}
where the domain $[-a,a]$ is the smallest set for which $\psi(z) > 0$.  \textcolor{black}{Physically, inequality \eqref{eq:inequality} implies an event in which we track a particle starting before it enters the laser and ending after it fully exits; the corresponding function $f(t)$ is padded on the left and right by zeros.\footnote{We assume that any constant offset or background has been subtracted from the fluorescence signal.}  Because $L_0$ is constant, the locations at which we start and stop tracking the particle are fixed.  Taking $t=0$ as the time when the particle is at the (arbitrary) ``center'' $z=0$ of the laser, $f(t)$ is defined on the interval $[-\tau,\tau]$, where $\tau = L_0/v$.}

\textcolor{black}{Equation \eqref{eq:modelreduction} implies that increasing or decreasing $v$ ``compresses'' or ``stretches'' the signal in the time domain.  Thus, we should be able to solve the reverse problem: estimate the relative velocities by undoing the deformations in such a way that the signals coincide.}  
To achieve this, it is convenient to consider Fourier representations of the signals.  
We denote the corresponding transforms of $\psi(z)$ and $[R^2 - z^2] \Theta[R^2 -z^2]$ by $\hat \psi(k)$ and $\hat g(k;R)$, where a simple computation yields\footnote{We use the normalization $\mathfrak F[\cdot]=\int_{-L_0}^{L_0} \cdot \exp(-ikz/L_0) {\rm d}z$, where $\mathfrak F[\cdot]$ is the Fourier transform acting on $\cdot$.}
\begin{align}
\hat g(k;R) \!&=\! \begin{cases}
\frac{-4RL_0^2 \cos(kR/L_0)}{k^2} + \frac{4L_0^3 \sin(kR/L_0)}{k^3} & \!k\ne 0 \\
\frac{4}{3}R^3 & \! k=0
\end{cases} \label{eq:chat}
\end{align}
and  $k = \pi n$ for any integer $n$.  In practice, we assume that $-M \le n \le M$, where $M$ is a mode-cutoff associated with the noise floor of the signal; see also Sec.\ \ref{sec:validation}.  Unless otherwise stated, all sums over $k$ range from $-M\pi$ to $M\pi$.   Note that $\hat \psi(k)$ is unknown.

In light of these assumptions, Eq.\ \eqref{eq:modelreduction} becomes
\begin{align}
f(t) &=  \frac{c \pi}{2L_0} \sum_k \hat \psi(k) \int_{-L_0}^{L_0} \!\!\!\!\dz \left[R^2 \!-\! (z\!-\!vt)^2 \right] \nonumber \\
&\qquad\qquad\qquad\qquad\Theta\left[R^2 \!-\! (z\!-\!vt)^2\right] e^{ikz/L_0}, \label{eq:fourierrep}
\end{align}
where we have used the fact that
\begin{align}
\psi(z) = \frac{1}{2L_0}\sum_k \hat \psi(k) e^{ikz/L_0}.
\end{align}
To further simplify this, use inequality \eqref{eq:inequality} to impose the periodicity relationship  $C(x,y,z-vt) = C(x,y,z-vt + 2nL_0)$, where $n\in \mathbb Z$ is an integer.  This assumption does not  change the signal on the domain $-L_0/v \le t \le L_0/v$.  However, it does allow us to invoke the convolution theorem \cite{Convolution}, which yields
\begin{align}
f(t) &= c \pi  \sum_k \hat \psi(k)\hat g(k;R)e^{ikt/\tau}, \label{eq:foft} 
\end{align}

Equations \eqref{eq:chat} and \eqref{eq:foft} illustrate how the measurand and measurement conditions affect the signal.  Varying the concentration $c$ and velocity $v$ (which controls $\tau$) alters the height and width of the signal.  Changes in particle size $R$ have more complicated effects as characterized by Eq.\ \eqref{eq:chat}, since a particle is convolved with a larger fraction of the laser profile as $R$ increases.  Taking the Fourier transform of Eq.\ \eqref{eq:foft} with respect to time (over the interval $-\tau$ to $\tau$) yields
\begin{align}
\hat f(k;\tau) = 2 c \pi \tau \hat \psi(k) \hat g(k;R).\label{eq:ft}
\end{align}
\textcolor{black}{The inclusion of $\tau$ in the argument of $\hat f(k;\tau)$ is to emphasize the time-domain over which the transform is taken, which is important in the following sections.}  Letting $f(t)$ and $f_0(t)$ denote test and reference signals, the laser profile $\hat \psi(k)$ is eliminated by taking the ratio
\begin{align}
\frac{v\hat f(k;\tau)}{v_0\hat f_0(k;\tau_0)} = \frac{c \hat g(k;R)}{c_0 \hat g(k;R_0)}, \label{eq:ratio}
\end{align}
where quantities with $0$ subscripts are associated with $f_0(t)$ and the trajectory length $L_0$ is assumed to be the same for both particles.  In the remainder of the manuscript, we assume that $\tau_0 = L_0/v_0$, i.e.\ the reference particle defines the length and time scales against which we compare all events. 

Equation \eqref{eq:ratio} is the key result of this section: it defines the shape of one signal in terms of another and as a function of the particle properties, provided the trajectories are identical.


\subsection{Converting between fixed-time and fixed-trajectory representations}
\label{subsec:transformations}

\textcolor{black}{In practice, it is difficult to measure an event over a fixed trajectory length and variable time domain.  It is more common for acquisition to occur in fixed time windows, which we assume to be $[-\tau_0,\tau_0]$.  However, if $\tau_0$ remains fixed, the corresponding trajectory lengths of each particles must vary, which violates the key assumption of the previous section, so that we cannot use Eq.\ \eqref{eq:ratio}.\footnote{Allowing trajectory length to vary per-particle would yield varying spectral representations of $\psi(z)$, making it difficult to eliminate the laser profile via a relationship akin to Eq.\ \eqref{eq:ratio}.}}

The resolution to this problem is to identify a mapping that converts a fixed-time Fourier representation to one on a fixed spatial domain.  \textcolor{black}{Observe first that in Eq.\ \eqref{eq:ft}, the dimensionless frequency $k$ is common to all signals; moreover, $\hat f(k;\tau)$ is linear in the particle-dependent time-scale $\tau=L_0/v$.  Instead of $\hat f(k;\tau)$ we are given a measured signal $\hat f_m(k;\tau_0)$, where $\tau_0 = L/v = L_0/v_0$ and $L\ne L_0$ if $v\ne v_0$.  The linearity of $\hat f(k;\tau)$ in $\tau$ suggests that there is a transformation matrix $\X$ such that } 
\begin{align}
\hat f(k;\tau) = \sum_{k'} X_{k,k'}(\tau,\tau_0)\hat f_m(k';\tau_0), \label{eq:lamtransform} 
\end{align}
where $\hat f(k;\tau)$ and $\hat f_m(k';\tau_0)$ are the constant-trajectory and constant-time representations of the event.  The subscript $m$ emphasizes that the latter is the measured signal.  The time domains associated with the right and left sides of Eq.\ \eqref{eq:lamtransform} are $D_v=[-\tau,\tau]$ and $D_0=[-\tau_0,\tau_0]$.  (The subscript $v$ on $D_v$ is to emphasize that $v$, not $L_0$, controls $\tau$.)    We temporarily assume that $\tau$, $\tau_0$, and $L_0$ are known; in the next section we show how to find these quantities.




Equation \eqref{eq:lamtransform} does not itself specify the operator $\X$.  Its definition depends on how we wish to extend or truncate a signal when mapping it to a constant-trajectory domain.  \textcolor{black}{There are two equivalent perspectives one may take:  active or passive.  In the former, we deform (i.e.\ expand or contract) the time-series so that its constant-trajectory domain $D_v$ becomes $D_0$.  This collapses all time-series onto one another.  In the latter perspective, we deform $D_0$ to become $D_v$, keeping the time-series unchanged.  While both approaches are equivalent, the corresponding derivations have subtle differences.}

To better understand these issues, consider the case in which  $v>1$.    Clearly $D_v$ is a subset of $D_0$.  In the passive approach, we are given $\hat f_m(k;\tau_0)$, i.e.\ the Fourier modes on $D_0$, and wish to find the corresponding modes on the smaller domain $D_v$.  In this sense, $\X$ is a ``domain-restriction'' operator.  Assume that inequality \eqref{eq:inequality} holds, so that $f_m(t) \to 0$ as $t\to \pm L_0/v$.  Defining the restriction $f_r(t) = f_m(t)$ for $t\in [-\tau,\tau]$, we take the Fourier transform of $f_r(t)$ on $D_v$ to find
\begin{align}
\hat f(k;\tau) &:= \hat f_r(k;\tau) \nonumber \\
 &= \frac{v_0}{v} \sum_{k'}{\rm sinc}\!\left( k \!-\! \frac{k'v_0}{v} \right) \hat f_m(k';\tau_0), \label{eq:embedding}
\end{align}
so that $X_{k,k'}:=(v_0/v){\rm sinc}(k-k'v_0/v)$ and the $:=$ symbol means that we {\it define} the left-hand side to be equal to the right-hand side.  We adopt the convention that
\begin{align}
{\rm sinc}(x) = \sin(x)/x.  
\end{align}
To arrive at this result via the active perspective, we scale the argument of $f_m(t)$ by a factor $v_0 / v$, so that $f_m(t) \to f_m(v_0 t /v)$.    Then, restricting $f_m(v_0 t /v)$ to the domain $D_0$ and taking the Fourier transform yields an estimate that we denote $\hat f_r(k;\tau_0)$ in a slight abuse of notation.\footnote{The restriction $f_r(t)$ is only defined on $D_v$, so that by the notation $\hat f(k;\tau_0)$ we mean the Fourier transform of the expanded version of $f_r(t)$, which has the constant-trajectory domain $D_0$.}  To convert this to $\hat f(k;\tau)$ (i.e.\ the mode-weight on the domain $D_v$), we invoke Eq.\ \eqref{eq:ratio} with $R=R_0$ and $c=c_0$, which yields $\hat f_r(k;\tau) = (v_0/v)\hat f_r(k;\tau_0)$.  This derivation suggests the interpretation that $\X$ is also a ``signal-expansion'' operator, since $v > v_0$.  We leave it as an exercise to the reader to show the the corresponding matrix $\X$ is identical to that given by Eq.\ \eqref{eq:embedding}.

When $v<1$, the domain $D_v=[-\tau,\tau]$ contains the interval $[-\tau_0,\tau_0]$, so that in the passive-interpretation, $\X$ is a ``domain-extension'' operator that  extends $f_m(t)$ from $D_0$ onto $D_v$.  While there is no unique extension $f_e(t)$, we assume without loss of generality that $f_m(t) \to 0$ as $t\to \pm \tau_0$.\footnote{Signals violating this condition can be treated as multiplets, which are considered in Sec.\ \ref{sec:applications}.}  This together with inequality \eqref{eq:inequality} suggests the definition
\begin{align}
f_e(t) := \begin{cases}
f_m(t) & t\in [-\tau_0,\tau_0] \\
0 & {\rm otherwise}.
\end{cases}
\end{align}
Again, taking the Fourier transform  on $D_v$, one finds
\begin{align}
\hat f(k;\tau) :=  \sum_{k'} {\rm sinc}\left(k\frac{v}{v_0} - k' \right)\hat f_m(k';\tau_0), \label{eq:extension}
\end{align}
so that $X_{k,k'}:= {\rm sinc}[k(v/v_0)-k']$.  In the active interpretation, $\X$ is a ``signal-contraction'' operator; we leave derivation of the equivalence with Eq.\ \eqref{eq:extension} as an exercise for the reader.    

Combining these results, we define $f(t)$ on $D_v$ to be
\begin{align}
f(t):=\begin{cases}
f_r(t) & v \ge v_0 \\
f_e(t) & v \le v_0,
\end{cases}\label{eq:fe}
\end{align}
with $\hat f(k;\tau)$ being the corresponding Fourier transform.  In both Eqs.\ \eqref{eq:embedding} and \eqref{eq:extension}, setting $v\to v_0$ yields the identity transformation, as expected. 

In practical settings, distinct signals $f_m(t)$ may not be identically centered on the domain $D_0$ (i.e.\ before applying $\X$).  However, Eq.\ \eqref{eq:modelreduction} assumes that the center of any given particle is at $z=0$ when $t=0$, and moreover $\X$ deforms the signal about $t=0$.  Thus, it is necessary to consider transformations of the form $f_m(t) \to f_m(t+\Delta t)$.  As before, assume that $f_m(t)$ is padded by zeros as $t \to \pm \tau_0$, so that cyclic permutations by sufficiently small $\Delta t$ only wrap zeros around the left and right sides of the peak.  Assuming that $f_m(t)$ is periodic with a period of $2\tau_0$, one finds that 
\begin{align}
f_m(t) &= \sum_k \hat f_m(k;\tau_0) e^{ikt/\tau_0} \nonumber \\  & \qquad \to \sum_k \hat f_m(k;\tau_0) e^{ikt/\tau_0 + ik\Delta t/\tau_0} \label{eq:roll}
\end{align}
so that $\hat f_m(k;\tau_0) \to \hat f_m(k;\tau_0)\exp\left(ik\Delta t/\tau_0\right)$ under time translation.  While the padding assumption is technically not necessary for the validity of Eq.\ \eqref{eq:roll}, it ensures that subsequent transformations by $\X$ do not truncate non-zero parts of the signal.  

\subsection{Signal Matching}
\label{subsec:matching}

Assume that we have reference and test signals $f_0(t)$ and $f_m(t)$, both mapped to the interval $-\tau_0 \le t \le \tau_0$ and sufficiently padded by zeros on left and right.  
Both Fourier expansions use the same set of frequencies.  Let $\k$ denote the vector of frequencies, and $\f_0$ and $\f_m$ denote the corresponding vectors of Fourier coefficients.  Also define the matrix operators $\X$, $\T$, and $\S$ having elements
\begin{align}
X_{k,k'}(v,v_0)&:=\begin{cases}
(v_0/v)\sinc(k-k'v_0/v) & v \ge v_0 \\
\,\sinc(kv/v_0 - k') & v \le v_0
\end{cases} \label{eq:expansionmat}\\
T_{k,k'}(\Delta t)&:=\delta_{k,k'}e^{ik\Delta t/\tau_0} \label{eq:rollmat} \\ 
S_{k,k'}(R;R_0) &:= \delta_{k,k'} \hat g(k;R_0)/\hat g(k;R) \label{eq:rscale}
\end{align}
where $\delta_{k,k'}=1$ if $k=k'$ and $\delta_{k,k'}=0$ if $k\ne k'$.  Equations \eqref{eq:expansionmat} and \eqref{eq:rollmat} derive from matrix analogues of Eqs.\ \eqref{eq:embedding}, \eqref{eq:extension}, and \eqref{eq:roll}.    The quantity $\hat g(k;R)$ is given by Eq.\ \eqref{eq:chat} and defines the relative transformation associated with changing particle radius.    The vector analogue of $\hat f(k;\tau)$ is denoted $\f(\k;\tau)$.

For noise-less signals and a known set of transformation parameters $\Delta t$, $R$, $R_0$, $v$, and $v_0$, Eq.\ \eqref{eq:ratio} implies that 
\begin{align}
&\lam \f_m(\k;\tau_0) - \f_0(\k;\tau_0) = 0 \label{eq:sigdiff}
\end{align}
where 
\begin{align}
\lam = \frac{(v/v_0)}{(c/c_0)}\S(R,R_0)\X(v,v_0)\T(\Delta t). \label{eq:lammat}
\end{align}
The interpretation of Eq.\ \eqref{eq:sigdiff} is straightforward.  The product $\lam \f_m(\k;\tau_0)$: (i) centers $f_m$; (ii) scales it to the domain $D_v$; (iii) matches the radius scaling to $R_0$; and (iv) normalizes the amplitude to the reference signal.  The order of operations of the matrices mirrors the assumptions of the analysis above and cannot be changed.\footnote{The transformations matrices do not all commute with one another.  For example, centering a peak and then truncating its edges can yield a different signal than truncating followed by centering. }  

In practice, the transformation parameters must be determined from noisy signals.  This motivates the objective
\begin{align}
\L=\Big|\Big|\lam \f_m - \f_0 \Big|\Big|^2 \label{eq:bigobj}
\end{align}
where the square is interpreted as the sum over magnitude-squared of the elements of the complex vector argument.  \textcolor{black}{Ostensibly minimizing $\L$ as a function of the $\Delta t$, $R$, $R_0$, $v$, $v_0$, $c$, and $c_0$ should yield their numerical values.  However, many of these quantities only appear in $\lam$ via the ratios $v/v_0$, $R_0/L_0$, and $c/c_0$.  As a result, we can only determine their relative values.  This reflects a deeper freedom that we have not yet exploited: the ability to arbitrarily pick the numerical scale of the coordinate system.  For convenience, we pick $c_0=1$, $L_0=1$ (i.e.\ the reference particle traverses a distance of $2$), and $v_0=1$, which implies that $\tau_0=1$, with all units now being dimensionless.  The latter two choices fix the time and length scales of the system, so that $R$, $R_0$, and $v$ are defined relative to the reference trajectory and velocity.  We use this convention throughout the remainder of the manuscript.}

With these choices, we define the true $c$, $v$, $R$, $R_0$, and $\Delta t$ to be solutions to the optimization problem
\begin{align}
\{c^*\!,v^*\!,R^*\!,R_0^*,\Delta t^*\} = \mathop{\rm argmin}_{\{c,v,R,R_0,\Delta t\}}\!\!\!\!\! \L(c,v,R,R_0,\Delta t;\f_m). \nonumber 
\end{align}
The objective $\L$ is highly non-linear and may have local minima.  Thus, it is important to identify reasonable initial points as inputs to the optimization.  While these issues are discussed at greater length in Sec.\ \ref{sec:validation}, we note that simple properties of the signal such as its height, width at half max, etc.\ yield initial conditions that should be sufficiently close to optimal solutions to ensure convergence of the optimization. 


\section{Validation with Experimental Data}
\label{sec:validation}

To validate the analysis presented in Sec.\ \ref{sec:theory}, we consider a collection of 1302 events measured in the optofluidic device described in Ref.\ \cite{Part1}.  The measurands are $15.3\,\mu$m diameter polystyrene microspheres with dispersed Dragon Green fluorophore.\footnote{Certain commercial equipment, instruments, software, or materials are identified in this paper in order to specify the experimental procedure adequately. Such identification is not intended to imply recommendation or endorsement by the National Institute of Standards and Technology, nor is it intended to imply that the materials or equipment identified are necessarily the best available for the purpose.}  We use a high-speed camera system to manually verify that all events correspond to singlets.  The fluorescence signals, denoted $\boldsymbol{\rm f}_m$, are discretized on a 16-bit data acquisition card that records samples at 2 MHz.  Thus, the $\boldsymbol{\rm f}_m$ are vectors whose entries correspond to the discrete time interval over which the data is collected.  The digitizer outputs fluorescence values in units of volts; we divide these measurements by $1$ V to non-dimensionlize.

As a preprocessing step, we use a moving average filter to smooth peaks followed by a polynomial fit peak finder to estimate the height $h_m$, width $w_m$, and the time interval of the location of the peak maxima for each event; see Ref.\ \cite{Part1,Kearsley1,Kearsley2}.  The first two quantities are used later in the optimization.  Using the third, we perform a cyclic permutation of the data (corresponding to temporal shifts) to approximately align the peak maximum with the midpoint of the time interval, which we rescale to be $[-1,1]$.  

Having estimated these quantities, we return to the original (non-smoothed) signal for subsequent analysis.  Because the signals contain noise (likely from the photodetector and/or counting statistics of photons), we perform a discrete Fourier transform (DFT) to identify a noise-floor, which begins around the tenth mode, corresponding to $M=9$ according to our indexing convention; see also Ref.\ \cite{Fourier1}.  We take this value to be a cutoff for a low pass filter and also keep the complex conjugate weights, which correspond to the last 9 modes (the case $k=0$ is its own conjugate mode).\footnote{The exact cutoff will depend on application and the noise-floor of the signal. }  Given that the DFT yields a spectral representation that is a continuous interpolation of the data, we can directly use the corresponding mode weights and frequencies in calculations involving Eq.\ \eqref{eq:bigobj}.  

To test Eq.\ \eqref{eq:bigobj}, we pick a reference event $\f_0$ at random from among those having a height and width roughly equal to the median value for this population.    The specific choice of this event is unimportant, since the goal is to demonstrate that all signals can be collapsed to any chosen reference.   Next, we choose $N_s$ additional samples $\f_m$ and define the modified objective
\begin{align}
\L_{N_s}&= \sum_{m=1}^{N_s} \L(c_m,v_m,R_m,R_0,\Delta t_m;\f_m) \label{eq:multiobjective}
\end{align}
where $R_0$ is the unknown radius of the reference particle and $(c_m,v_m,R_m,\Delta t_m)$ are the unknown transformation parameters associated with $\f_m$.  In Eq.\ \eqref{eq:multiobjective}, $m$ has been upgraded from a subscript to an index.  Because the beads have on the order of $10^9$ fluorophores per particle, we anticipate that variations in $c_m$ (e.g.\ due to shot noise) are negligible.  Thus, we set $c_m=1$ for all values of $m$, so that it does not play a further role in the analysis.

In principle, minimizing $\L_{N_s}$ with respect to the unknown parameters yields estimates of their values, as well as a ``consensus'' estimate of $R_0$.  However, several computational considerations limit our ability to use Eq.\ \eqref{eq:multiobjective} directly.  First, the objective function is nonlinear and contains on the order of $20N_s$ terms and $3N_s$ parameters.  Second, the objective has rapid oscillations due to the forms of $\X(v)$ and $g(k;R)$.  Third, in the limit that $R\to 0$ for $R/R_0$ constant, the ratio $g(k;R_0)/g(k;R) \to (R_0/R)^3$, so that $\L_{N_s}$ develops a connected set of solutions.  That is, only the ratio of $R_0$ to $R$ can be estimated from $\L_{N_s}$.

A resolution to these problems is to consider a regularized version of the objective
\begin{align}
\L_r  = \L_{N_s} + \epsilon (R_0 - \bar R_0)^2, \label{eq:regobjective}
\end{align}
where $\epsilon \ll 1$ is a regularization parameter and $\bar R_0$ is an estimate of the particle radius given by outside sources of information.  For the beads under consideration, manufacturer specifications indicate that the average radius is approximately $7.625$ $\mu$m, which we use as the value for $\bar R_0$ in dimensional units.  To convert to dimensionless units, we note that: (i) the digitizer samples at 2 MHz; (ii) each peak is 1200 samples long; and (iii) the average velocity as estimated from time-of-flight between two interrogation regions is 337 mm/s.  Thus, the characteristic distance $d_c$ traversed by a particle during an event is 
\begin{align}
d_c = \frac{1200\, {\rm samples}}{2 {\rm MHz}}337 {\rm mm/s} \approx 202\,\mu{\rm m}
\end{align}
The characteristic radius 7.625 $\mu$m normalized by $d_c$ and doubled (to account for mapping to $-1\le z \le 1$) yields $\bar R_0 = 0.0755$.   We also set $\epsilon = 10^{-2}$ and fix $N_s=5$.

\begin{figure}
\includegraphics[width=7.5cm]{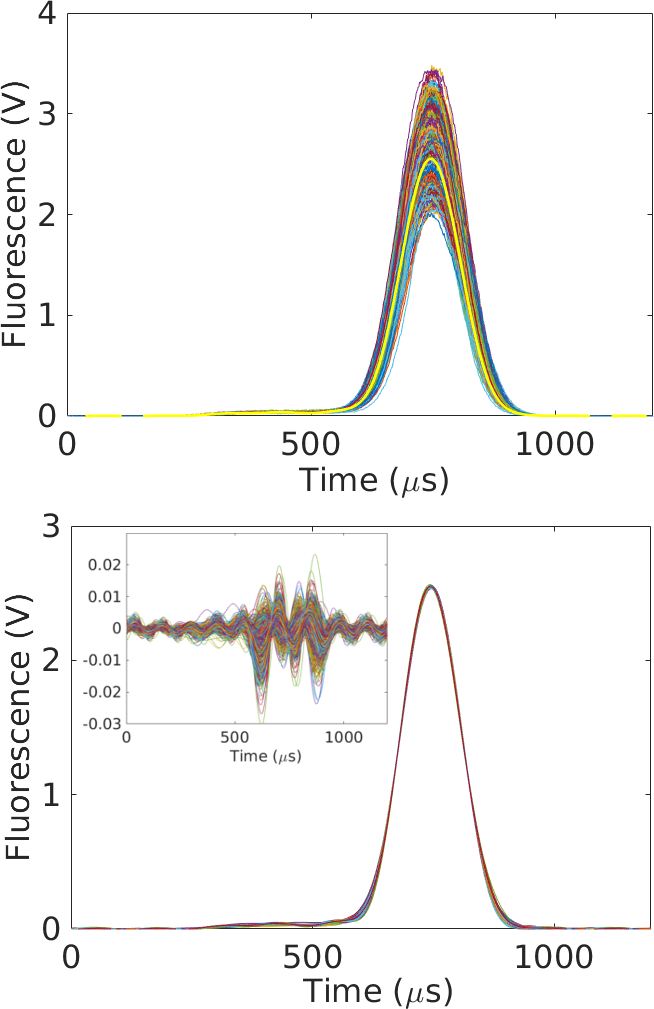}\caption{Example of data collapse for 1302 time-series.  {\it Top:}  The original curves to which we apply the scale transformations described in the main text.  The bold-yellow time-trace is used as the reference.  The time and fluorescence values are given in their original units, although subsequent analysis is done after non-dimensionalizing.  {\it Bottom:}  All 1302 curves after data collapse and conversion back to the original units.  To achieve, collapse, we use Eq.\ \eqref{eq:sigdiff} to turn each measured curve into a realization of $f_0(t)$.  For the purposes of defining units in this and other figures, we adopt this active interpretation of the transformations. The inset shows the point-wise residuals between the transformed curves and reference.  The difference is normalized by the maximum value of the reference.  \textcolor{black}{Note that the residuals have the characteristic frequency of the mode cutoff.}}\label{fig:collapses}
\end{figure}

\begin{figure}
\includegraphics[width=7.5cm]{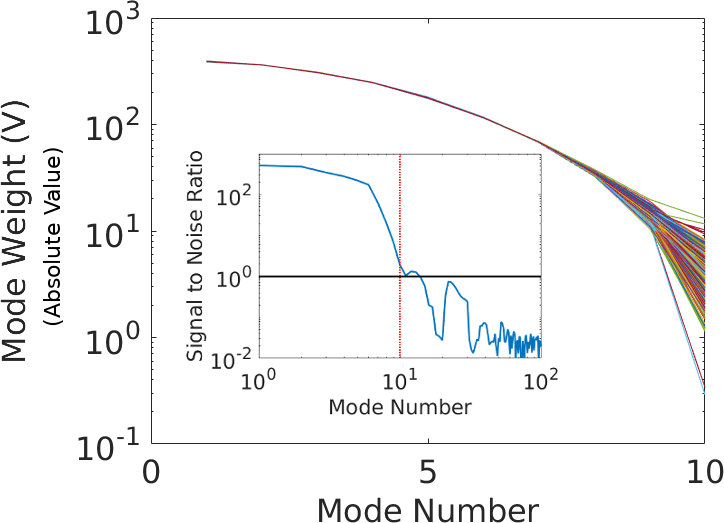}\caption{Absolute values of the mode-weights associated with the 1302 curves in Fig.\ \ref{fig:collapses} after transformation and collapse via Eq.\ \eqref{eq:sigdiff}.  The inset shows the signal-to-noise ratio, which was computed by: (i) using the first 10 modes to determine the transformation parameters; (ii) applying the transformation matrices to the first 100 modes of each signal; and (iii) computing the ratio of the absolute value of the mean to standard deviation of the resulting mode-weights.  The dotted-red vertical line shows the mode cutoff, which occurs when the mode-weights approach a signal-to-noise ratio of unity.  }\label{fig:collapsed_modes}
\end{figure}

To estimate transformation parameters associated with the remaining curves, we minimize $\L$ for each of the remaining 1296 curves separately, using the consensus value of $R_0$.  The results of this exercise are shown in Fig.\ \ref{fig:collapses}.  The top subplot shows the original time-traces, while the bottom subplot shows the data collapse; the inset shows the relative errors.  See also Fig.\ \ref{fig:collapsed_modes}.  We also compute the sample mean $\bar R$ and sample variance $\sigma_R^2$ estimates of the particle radius to estimate the coefficient of variation (CV) given by $CV_R=\sigma_R/\bar R$ \cite{GUM}.  We find $CV_R = 3.18$~\%,  which is consistent with the manufacturer specified range of particle sizes ($4.4$\%) and the number of measurements considered. 

\begin{figure}
\includegraphics[width=7.5cm]{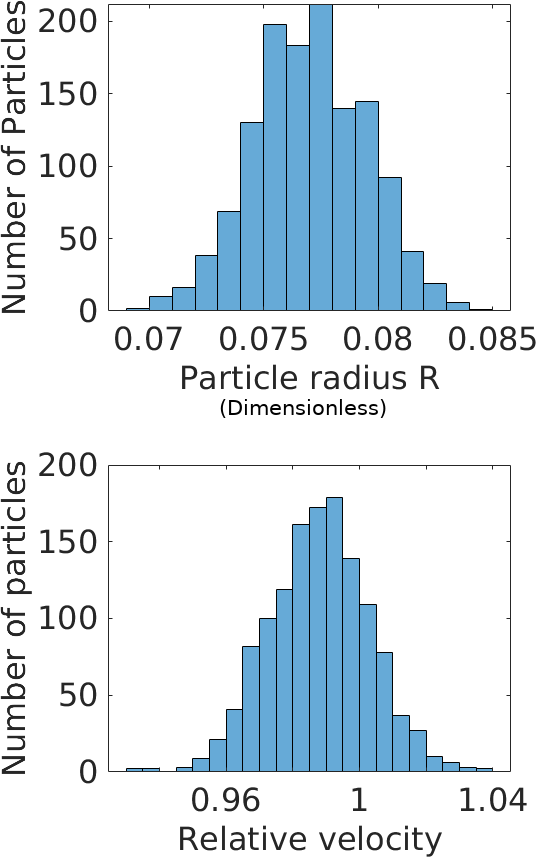}\caption{Histograms of relative particle sizes (top) and velocities (bottom).  The coefficient of variation (CV) in particle radii is 3.18\%, which matches well with the manufacturer specification of 4.4\% (Bangs Laboratories, FSDG009).}\label{fig:velhist}
\end{figure}

\section{Applications}
\label{sec:applications}

We now consider two applications of the analysis discussed in Sec.\ \ref{sec:validation}: estimating per-event measurement reproducibility and doublet deconvolution.  A key theme of these examples is that UQ enables one to extract otherwise inaccessible information from cytometry measurements.  In particular, UQ allows us to resolve a {\it uniqueness problem} of identifying the most probable singlets that generate a multiplet.  Our main task is to recast the previous results in a probabilistic framework and demonstrate how this leads to new kinds of signals analyses.

\subsection{Uncertainty Quantification}
\label{subsec:UQ}

\begin{figure}
\includegraphics[width=7.5cm]{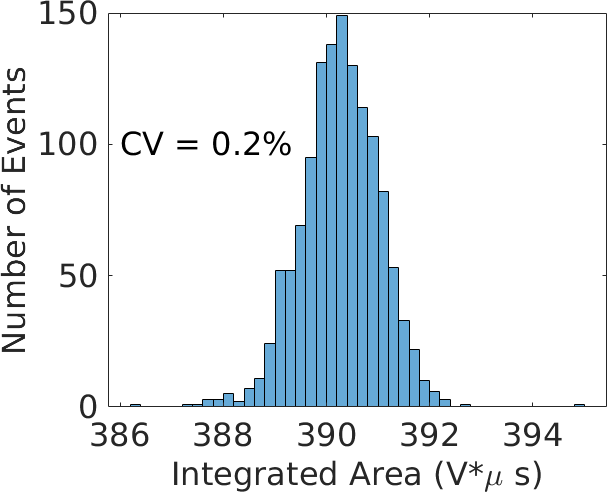}\caption{Histogram of integrated areas of the transformed signals in Fig.\ \ref{fig:collapses}.}\label{fig:relvarhist}
\end{figure}

Equation \eqref{eq:sigdiff} implies that the optimal transformation parameters undo signal deformation due to effects such as particle speed, size, brightness, etc.  Thus, each transformed signal $\lam \f_m$ can be viewed as a realization of a measurement of identical particles, i.e.\ a version of $\f_0$.  Treating these as independent and identically distributed  random variables, we can estimate the per-event reproducibility in terms of corresponding statistical estimators.  For example, if $A_j$ denotes the integrated area of the $j$th transformed signal, then reproducibility in area measurements is given in terms of a standard deviation $\sigma_A$ computed via
\begin{align}
\sigma_A^2 &= \frac{1}{N_e-1}\sum_{j=1}^{N_e} (A_j - \bar A)^2 \label{eq:sampvar} \\
\bar A &= \frac{1}{N_e} \sum_{j=1}^{N_e} A_j \label{eq:sampmean}
\end{align}
where $N_e$ is the number of events.\footnote{In Sec.\ \ref{sec:theory} we use the subscript $m$ to denote a measured time-series, which we implicitly index from $1\le m \le N_e$.  Unless otherwise specified, in this section we use the subscript $j$ when referring to transformed signals.  Note that the number of transformed signals is still $N_e$.}  Figure \ref{fig:relvarhist} shows the results of this analysis for the 1302 curves considered in Fig.\ \ref{fig:collapses}.
  The inset to the bottom plot of Fig.\ \ref{fig:collapses} indicates that the point-wise (in time and relative to the reference amplitude) reproducibility in shape of the filtered signals is on the order of 2 \% or less.  

We make this last observation more precise by considering the Fourier modes of each realization of $\f_0$, which we denote by $\f_{0,j}$.  As a practical matter, it is easier to consider the real and imaginary parts of $\hat f_{0,j}(k;\tau_0)$ separately, since their fluctuations should be independent.  Dropping explicit reference to $\tau_0$, we decompose $\hat f_{0,j}(k) = \hat f_{j,{\rm Re}}(k) + i\hat f_{j,{\rm Im}}(k)$, where
\begin{subequations}
\begin{align}
{\rm Re}[ \hat f_{0,j}(k)] = \hat f_{j,{\rm Re}}(k) = \bar f_{\rm Re}(k) + \Delta \hat f_{j,{\rm Re}}(k) \\
{\rm Im}[ \hat f_{0,j}(k)] = \hat f_{j,{\rm Im}}(k) = \bar f_{\rm Im}(k) + \Delta \hat f_{j,{\rm Im}}(k), 
\end{align}
\end{subequations}
$\bar f_{{\rm Re}}(k)$ [$\bar f_{{\rm Im}}(k)$] are average real [imaginary] Fourier modes, and $\Delta \hat f_{j,{\rm Re}}(k)$ $[\Delta \hat f_{j,{\rm Im}}(k)]$ are random variables associated with uncertainty in each mode.  The average mode weights are constructed by analogy to Eq.\ \eqref{eq:sampmean}.  As a bookkeeping step, it is useful to define a vector $\hat {\mathfrak f}_j$ whose first $M+1$ elements (where $M$ is the mode cutoff) are the real parts of modes $k=0,\pi,2\pi,...,M\pi$ of $\f_{0,j}$ and whose next $M$ elements are the corresponding imaginary parts for $k\ne 0$; see Appendix \ref{app:opt} for more detailed motivation of this decomposition.\footnote{Note that $\hat {\mathfrak f}$ contains all of the same information as $\f$ because $f(t)$ is real.}  There are  $N_e$ such realizations of $\hat {\mathfrak f}_j$, where $1\le j \le N_e$.  The corresponding average (sample mean) weights and perturbations are $\bar {\mathfrak f}$ and $\Delta \hat {\mathfrak f}_j$.  We also use the $\hat {\mathfrak f}_j$ to construct a covariance matrix $\Xi$, as well as the probability of a deviation $\Delta \hat {\mathfrak f}$ 
\begin{align}
P_1(\Delta \hat {\mathfrak f}) \propto \exp\left[-\frac{1}{2} \Delta \hat {\mathfrak f}^{\rm \,T} \,\Xi^{-1}   \Delta \hat {\mathfrak f} \right],  \label{eq:singprob}
\end{align}
where the proportionality factor depends on the dimensionality and determinant of $\Xi$.  Equation \eqref{eq:singprob} characterizes uncertainty in the shape of a signal.\footnote{The real basis underlying $\hat {\mathfrak f}$ in Eq.\ \eqref{eq:singprob} allows us to express correlations between modes in terms of the usual covariance matrix.}  Likewise, we construct a probability density 
\begin{align}
Q_1(\Delta \chi) \propto \exp\left[-\frac{1}{2} \Delta \chi^{\rm \, T} \Upsilon^{-1} \Delta \chi \right]
\end{align}
where $\Delta \chi^{\rm \, T}=(R-\bar R, v - \bar v)$ characterizes deviation from the average radius and velocity, and $\Upsilon$ is the corresponding covariance matrix.  In the next section, we show how these probability densities play a fundamental role in the tasks of doublet  deconvolution.

\subsection{Doublet Deconvolution}
\label{subsec:doublet}

As an illustration of how UQ can inform downstream analysis, consider the task of doublet deconvolution.  In typical cytometry protocols, such data is  identified indirectly via gating procedures and subsequently rejected.  We propose an alternate  strategy based on constrained optimization.

First assume that a doublet $d(t)$ is a linear combination singlet signals.  In Fourier space this implies
\begin{align}
\hat {\mathfrak d} &= \boldsymbol \Theta^{-1}(c_1,v_1,R_1,R_0,\Delta t_1) [\bar {\mathfrak f} + \Delta \hat {\mathfrak f}_1] \nonumber \\ &\qquad + \boldsymbol \Theta^{-1}(c_2,v_2,R_2,R_0,\Delta t_2) [\bar {\mathfrak f} + \Delta \hat {\mathfrak f}_2], \label{eq:doublet}
\end{align}
where the parameters $c_j$, $v_j$, $R_j$, and $\Delta t_j$ transform the reference signal into the corresponding singlets that comprise the doublet, and $\hat {\mathfrak d}$ and $\boldsymbol \Theta$ are the representations of $\hat \d$ and $\lam$ in the same basis as $\Delta \hat {\mathfrak f}$ (cf.\ the Appendix).  We use the inverse $\boldsymbol \Theta^{-1}$, since we are transforming {\it from} the reference signal to the measured signal.    Generalizing Eq.\ \eqref{eq:doublet} to an arbitrary multiplet $m(t)$, yields
\begin{align}
\hat {\mathfrak { m}} = \sum_j \boldsymbol \Theta^{-1}(c_j,v_j,R_j,R_0,\Delta t_j) [\bar {\mathfrak f} + \Delta \hat {\mathfrak f}_j]. \label{eq:multconstraint}
\end{align}

Assuming that a sample is known to be a multiplet comprised of $\mathcal M$ singlets, the probability that the singlets have a given set of deformations $\Delta \mathfrak f_1,\Delta \mathfrak f_2,...,\Delta \mathfrak f_{\mathcal M}$ {\it before transformation to $\hat {\mathfrak {m}}$} can be expressed as
\begin{align}
P_{\mathcal M}(\Delta \hat {\mathfrak f}_1,\Delta \hat {\mathfrak f}_2,...,\Delta \hat {\mathfrak f}_{\mathcal M}) = \prod_{j=1}^{\mathcal M} P_1(\Delta \hat {\mathfrak f}_j). \label{eq:multprob}
\end{align}
Likewise, the probability of a set of transformation parameters is given by
\begin{align}
Q_{\mathcal M}(\Delta \chi_1, \Delta \chi_2,...,\Delta \chi_{\mathcal M}) = \prod_{j=1}^{\mathcal M} Q_1(\Delta \chi_j) \label{eq:transprobs}
\end{align}
We assume that the true values of $\Delta \hat {\mathfrak f}_j$ and $\Delta \chi_j$ are those that maximize the joint probability $P_{\mathcal M} Q_{\mathcal M}$. However, the singlets must reconstruct the multiplet exactly.  Together this suggests the objective
\begin{align}
\L_{\mathcal M} = \sum_j \Delta \hat {\mathfrak f}_j^{\rm \,T}{\boldsymbol \Xi}^{-1}\Delta \hat {\mathfrak f}_j  + \Delta \chi_j^{\rm \T}\Upsilon^{-1} \Delta \chi_j \label{eq:doubobj}
\end{align}
which we minimize with respect to the $\Delta \hat {\mathfrak f}_j$ and $\Delta \chi_j$, subject to the constraint given by Eq.\ \eqref{eq:multconstraint}.  

This optimization problem is performed over $\mathcal M(2M+1) + 4\mathcal M$ parameters, where $\mathcal M$ is the number of singlets comprising the signal, and $M$ is the mode cutoff.  In practice, however, this can be reduced to a $4\mathcal M$ dimensional problem, since the constraint is linear in the $\Delta \hat {\mathfrak f}_j$.  We leave details of the optimization to the Appendix.  We also set $c_j=1$, consistent with Sec.\ \ref{sec:validation}.

\begin{figure}
\includegraphics[width=7.5cm]{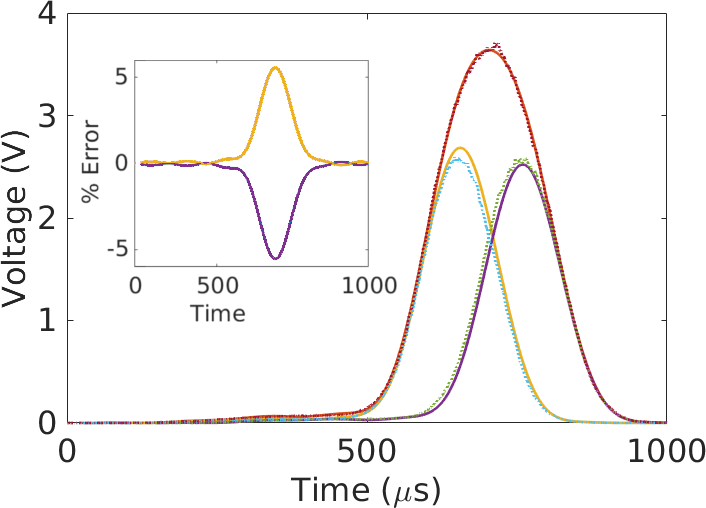}\caption{Example of doublet deconvolution.  Signals associated with two known singlets were added together to make a synthetic doublet.  Original signals are shown in dotted lines.  The solid lines show the recovered signals after deconvolution (purple and yellow) as well as the filtered doublet (red).  The inset shows the error in the recovered singlets relative to the original filtered singlets, normalized by the maximum values of the latter.}\label{fig:doubdecon}
\end{figure}

\begin{figure}
\includegraphics[width=7.5cm]{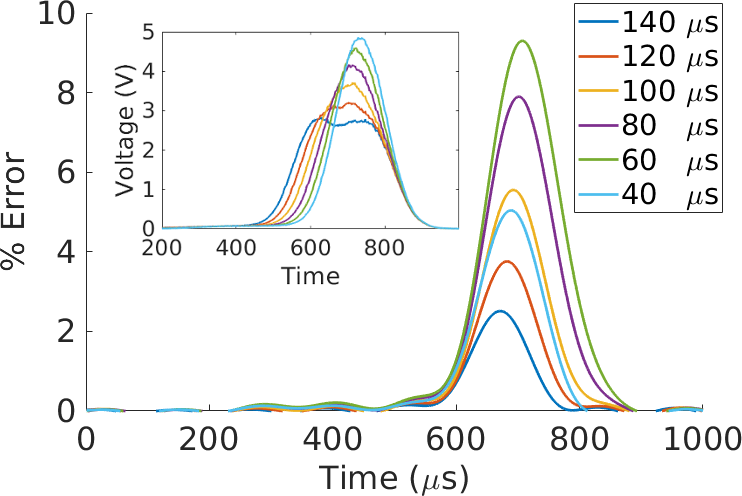}\caption{Impact of peak separation on doublet deconvolution.  The inset shows synthetic doublets constructed by varying the peak separation distance of the singlets in Fig.\ \ref{fig:doubdecon}.  The main figure shows the time-dependent relative difference between one of the recovered singlets and its true (filtered) signal.  Colors have the same interpretation in both the main plot and inset.  }\label{fig:manyerrors}
\end{figure}

\begin{figure}
\includegraphics[width=7.5cm]{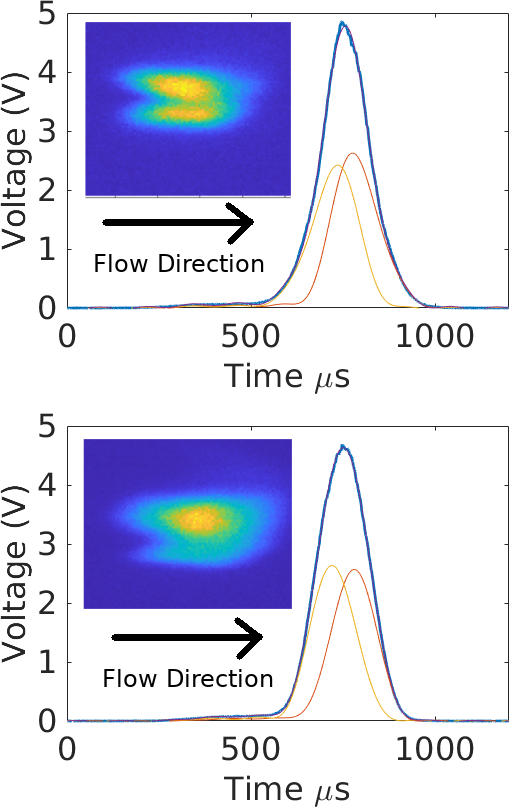}\caption{Deconvolution of two visually verified doublets.  The inset shows false color images of the particles traversing the laser interrogation region.  In the inset, the horizontal axis is parallel to the flow direction, and the vertical direction is parallel to the laser.  The horizontal elongation is due to blurring.  In both plots, two lobes are visible.  The main figure shows the original signal (blue), as well as the reconstructed singlets (orange and yellow).  The sum of singlets is superimposed in purple over the blue curve but is visually indistinguishable.  In the bottom plot, one particle is likely rotated out of the focal plane, which would account for the differences in visual intensities relative to signal peaks.  \textcolor{black}{This assumes the geometric factors associated with illumination and light collection are constant for both particles; see Ref.\ \cite{Part1} for the validity of this assumption.}} \label{fig:realdoublets}
\end{figure}

Figure \ref{fig:doubdecon} illustrates the results of this analysis applied to a synthetic doublet constructed from two singlets taken from Fig.\ \ref{fig:collapses}.  The analysis recovers the original signals to within roughly 5\% accuracy or better point-wise in time.  The actual transformation parameters associated with each particle are recovered to within approximately 2\%.  Figure \ref{fig:manyerrors} shows the relative errors (with doublets inset) associated with different peak spacings between the singlets.  All reconstructions have point-wise errors of less than 10\%, even for closely spaced singlets.  

Figure \ref{fig:realdoublets} shows the results of this analysis applied to two measured signals that were visually verified to be doublets.  In the second event (bottom plot), one of the beads was likely out-of-focus of the camera, leading to a reduced visual signal.  See Ref.\ \cite{Part1} for more details.  In both cases the analysis recovers singlets that are consistent with the synthetic examples shown in Fig.\ \ref{fig:doubdecon}.  While more work is needed to make this analysis fully robust for commercial settings, the examples provided herein indicate that the constrained optimization formulation is a useful tool for doublet deconvolution.

\section{Discussion}
\label{sec:discussion}

\subsection{The Role of Modeling in Cytometry}
\label{subsec:modelincyto}

From a metrology perspective, the operation of a cytometer is at odds with the fundamental assumptions used to characterize reproducibility and uncertainty.  The inherent separation of scales -- thousands of distinct, micron-sized cells traveling meters per second in a complicated fluid-dynamic system -- coupled with the native variability of biological systems means that it is challenging to repeat an independent measurement on the same particle in the same optical region.  Moreover, as Fig.\ \ref{fig:collapses} demonstrates, even small variations in {\it reference materials} can lead to dramatic changes in raw measurement signals.  Thus, a key challenge in characterizing the accuracy of a cytometer arises from an inability to design operating conditions that isolate individual sources of uncertainty.  In other words, experimental {\it analysis} of uncertainty is difficult.

The modeling herein provides a distinct approach to this problem through {\it synthesis, i.e.\ building up a prediction of the experimental result,} taking into account the cumulative effects of multiple sources of variation.  Under ideal circumstances, comparing this synthetic result with reality leads to a unique quantification of the physical phenomena (e.g.\ particle size, speed, etc.) that generate each signal.  Any remaining variation that we cannot account for is then treated as a reasonable proxy for reproducibility.  

This distinction between analysis and synthesis highlights both the potential roles and challenges of using mathematical modeling for device characterization.  In particular, the model permits us to infer the relative magnitudes of coupled physical effects, thereby compensating for experimental limitations.  But critically, the accuracy of these estimates relies on the validity of the underlying theoretical assumptions.  For example, the assumption that the particles are spherical may not be appropriate for deformable cells, requiring revision of Eq.\ \eqref{eq:fourierrep}.  Likewise, the properties of laser profiles considered herein may not apply to all cytometers.  

These observations suggest a need for deeper coordination between UQ, signals analysis, and design of cytometers.  While models can often be revised to account for increasingly complex phenomena (e.g.\ deformable cells, non-uniform lasers), computations invariably become too expensive to be useful.  In such cases, theory can instead inform design changes that may be experimentally achievable and lead to improved accuracy through consistency with modeling assumptions.  See also Ref.\ \cite{Part1} for related ideas.

\subsection{Further Applications and Open Directions}

The analysis presented herein offers routes to solving several outstanding problems in signals analysis for cytometry.  In particular, the ability to quantify reproducibility of a measurement suggests a task wherein one seeks to {\it minimize} this uncertainty as a function of operating conditions, e.g.\ flow velocity and degree of focusing, particle density, etc.  Moreover, reproducibility estimates suggest the possibility for propagating uncertainty into populations studies so as to inform best gating and classification strategies.  In this spirit, the work presented in Ref.\ \cite{Antibody} may be relevant.

Recent publications have also suggested the importance of cell deformability during the measurement process.  Equation \eqref{eq:modelreduction} makes a key simplifying assumption that requires modification in this case.  To the extent that it is possible to parameterize deformation modes (e.g.\ in terms of spherical harmonics or empirically defined shapes), the transformations associated with Eq.\ \eqref{eq:chat} can be generalized to account for non-spherical particles.  
This ability to extract shape information from a relatively simple time-series could yield significant improvements in throughput relative to imaging cytometers while still characterizing more nuanced information about the cell status.

\begin{figure*}
\includegraphics[width=\textwidth]{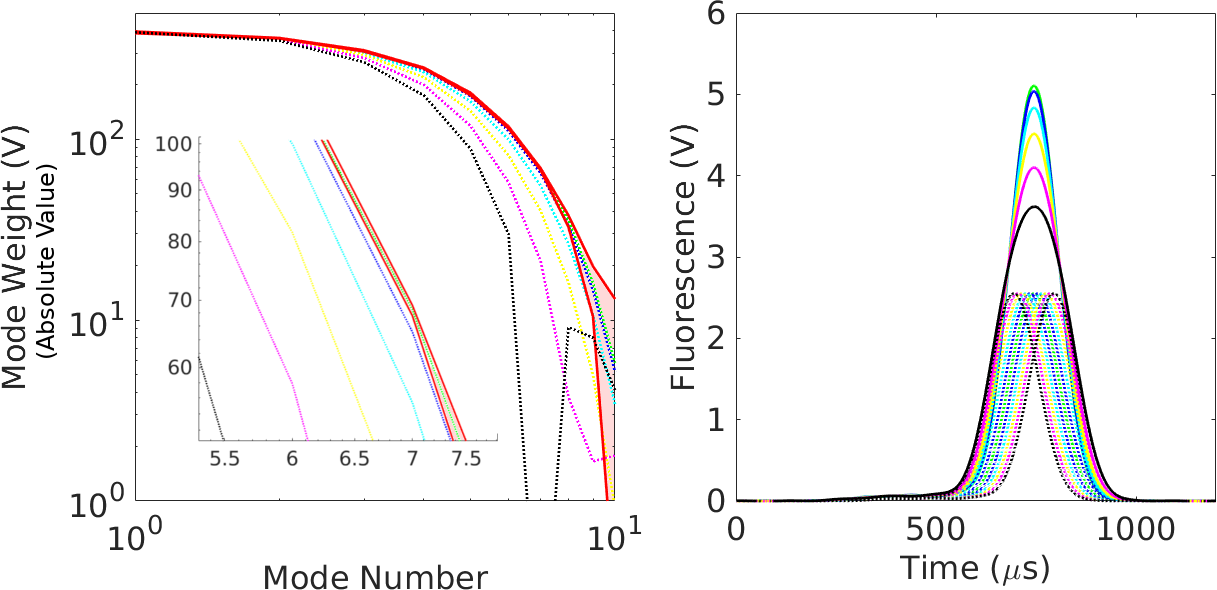}\caption{{\it Left:} Fourier spectra of synthetic doublets (dotted) compared with the uncertainty window for singlets (solid red).  For a fixed mode number, the uncertainty window for singlets is the set of all mode weights between the minimum and maximum values computed in Fig.\ \ref{fig:collapses}.  Synthetic doublets are composed of two copies of a singlet (taken at random from Fig.\ \ref{fig:collapses}) whose peaks are separated by 0 (green), 10 (blue), 20 (cyan), 30 (yellow), 40 (magenta), and 50 (black) microseconds.  The doublet mode weights are divided by 2 to normalize the them to scale of the singlets. Note that after a separation of only 10 microseconds the doublet spectra fall outside the uncertainty window of the singlets.  {\it Right:}  The doublets whose spectra are shown on the left plot.  Solid lines are the doublets, whereas dotted lines of the same color are the corresponding singlets.  The colors have the same interpretation as on the left plot.  Note that all doublets appear as a single peak and are visually difficult to distinguish from singlets. }\label{fig:doubletspectra}
\end{figure*}

The problem of {\it doublet identification} is also largely unresolved.  Traditional strategies address this task through subjective gating of populations.  However, the distribution of Fourier spectra may provide more objective, probability-based methods.  The left plot of Fig. \ref{fig:doubletspectra} shows the envelope of spectra (red region) associated with the distribution of transformed singlets in Fig.\ \ref{fig:collapses}.  The right plot shows a collection of synthetic doublets; their corresponding spectra (normalized to the same area as the mean singlet) are shown relative to the singlet envelope on the left.  Note that for almost any peak separation at least one mode associated with the doublet falls outside the admissible window for singlets.  This suggests that signal {\it shape} may be an exquisitely sensitive tool for identifying events that are candidates for our doublet deconvolution algorithm.  Such questions, however, are left for future work.  

\subsection{Metrics in the Context of Past Work}

Signals analysis has been integral to cytometry since its inception \cite{Sigs1,Sigs2,Sigs3,Sigs4,Fourier1}.  However, virtually all techniques characterize cells in terms of properties such as the signal area, height, width, etc., and the justifications for such analyses are not universally valid.  For example, common practice dictates that forward-scatter (FSC) vs area or FSC-height measurements are appropriate for detecting doublets, as cells should nominally follow one another single file \cite{Doublets1}.  But as  Fig.\ \ref{fig:realdoublets} illustrates, cells may pass the laser interrogation region side-by-side or even at a diagonal to the flow direction. A recent study of inertial effects also suggests that typical flow focusing strategies are ineffective \cite{Part1}.  Furthermore, biological processes of interest may interfere with measurements typically used to distinguish doublets \cite{DoubletProb1}.  

Many of these problems arise from the fact that quantities such as signal area, width, and height are functionals of high-dimensional data (i.e.\ the full event time-trace) that produce a scalar.  Thus, much of the underlying information that could be used to characterize measurands is lost before signals are actually compared.  This suggests a need to revisit the order of operations, e.g.\ by directly comparing full signals {\it before} quantifying their differences in terms of simple descriptors.

To better understand this point, it is useful to consider the concept of a {\it metric}, which addresses the question:  how ``far apart'' are the generic objects $h$ and $g$?  In other words, the metric, often denoted $d(h,g)$, defines a notion of distance appropriate to the structure of $h$ and $g$.   While a complete treatment is beyond the scope of this work (see Ref.\ \cite{Kreysig}), we note two fundamental properties: (i) $d(h,g) \ge 0$ for any objects $h$ and $g$, i.e.\ all distances are non-negative; and (ii) $d(h,g) = 0$ implies $h=g$.

While seemingly abstract, these observations have important ramifications for UQ of cytometry.  Conventional analyses implicitly use the absolute value metric
\begin{align}
d(\mathcal P[f_j(t)],\mathcal P[f_k(t)])=|\mathcal P[f_j(t)]-\mathcal P[f_k(t)]| \label{eq:absmetric}
\end{align}
and related notions of Euclidean distance as the basis for comparing measurements, where $\mathcal P[f_j(t)]$ is a functional that returns a scalar value (e.g.\ area, height) associated with the time-series $f_j(t)$.  Since the dimensionality of $\mathcal P[f_j(t)]$ is low, the statement $d(h,g)=0$ applied to Eq.\ \eqref{eq:absmetric} implies that two distinct time-series may still be treated as  equivalent.   For example, a large, dim cell may yield the same integrated area as a small bright one, despite the objects being very different.  Yet according to Eq.\ \eqref{eq:absmetric} both objects would be the same based on intensity alone.  Moreover, such approaches make it impossible to separate effects (e.g. flow rate) that may change the signal shape while keeping scalar properties such as the height constant.  Such shortcomings have limited UQ studies to those effects due to photodetectors and total uncertainties as characterized by population histograms \cite{UQ1,UQ2,UQ3,UQ4}.

The objectives considered herein address such problems by defining the metric in terms of the time-series explicitly; that is, Eq.\ \eqref{eq:bigobj} considers the situation $d(\f_j,\f_k) = ||\f_j - \f_k ||$ in which the notion of distance $||\cdot||$ is applied directly to the full time-series (expressed as its Fourier transform).  Equation \eqref{eq:bigobj} in particular is a variation on the commonly used $\ell_2$ or ``sum-of-squares'' metric \cite{Kreysig}.  A benefit of this approach is that it leverages the full information content of each signal to characterize multiple sources of variation.  However, the resulting analysis is more computationally expensive and does not have a simple interpretation.  Nonetheless, the example of doublet deconvolution highlights the usefulness of such techniques, and we speculate that more advanced signals analyses in cytometry will require further development of appropriate metrics.

\subsection{Limitations}

The modeling framework described Sec.\ \ref{subsec:model} sets forth the minimum assumptions required for the validity of our analysis.  A key goal of the theory is to avoid the need for detailed device characterization, which can be costly.  However, our main assumptions on the laser profile and geometric factors may not be applicable to all cytometers.  In such cases, using our analysis would decrease confidence in results by introducing additional model-form uncertainty.

In principle, such limitations can be overcome by more detailed modeling of the experimental system.  The general structure of Eq.\ \eqref{eq:bigobj} can be maintained, although the transformation matrix $\lam$ may not exist in closed form.  Rather, it may be necessary to evaluate it on-the-fly in terms of a simulation or other computational model of the measurement process.  In this case, a key challenge will be to formulate an optimization routine that can minimize Eq.\ \eqref{eq:bigobj} in a reasonable amount of time.  Reduced-order modeling or computationally inexpensive approximations are possible routes for addressing such problems.

\appendix

\section{Optimization for Multiplet Deconvolution}
\label{app:opt}

In this appendix we describe the mathematical formulation and solution of optimization for multiplet deconvolution.  For convenience, we restate the key equations underlying this problem.  Specifically, the objective is given by
\begin{align}
\mathcal L_d &= \sum_{j=1}^{\mathcal M} \Delta \hat {\mathfrak f}_j^{\rm T} \,\, \Xi^{-1} \, \Delta \hat {\mathfrak f}_j  + \Delta \chi_j^{\rm \T}\Upsilon^{-1} \Delta \chi_j\label{eq:appdoubobj} 
\end{align}
whereas the constraint in the complex basis is
\begin{align}
\hat {\boldsymbol {\rm m}} &= \sum_j \lam^{-1}(c_j,v_j,R_j,\Delta t_j) [\bar {\boldsymbol{\rm f}} + \Delta \f_j], \label{eq:appmultconstraint} 
\end{align} 
where $\bar {\boldsymbol{\rm f}}$ is the complex representation of $\bar {\mathfrak f}$.  Our goal is to convert Eq.\ \eqref{eq:appmultconstraint} to the same basis as $\hat {\mathfrak f}$ and then simplify the optimization problem implied by Eq.\ \eqref{eq:appdoubobj}.

Define $\lam_j^{-1}= \lam^{-1}(c_j,v_j,R_j,\Delta t_j)$.  Next, note that a vector $\f$ has $2M+1$ elements.  Because we compute them  via a DFT, we take the convention that the first $M+1$ modes correspond to $k=0,\pi,2\pi,...,M\pi$, whereas the last $M$ modes correspond to $k=-M\pi,-(M-1)\pi,...,-\pi$.  Moreover, because the signals are real in the time-domain, we know that $\hat f(k) = \hat f(-k)^\star$, where $\star$ denotes the complex conjugate.  This implies that all of the relevant information about $\f$ is contained in $\hat {\mathfrak f}$, since only $M$ complex Fourier modes are needed to describe the signal.  To see this explicitly, express an arbitrary $\f$ as
\begin{align}
\f &= \begin{pmatrix}
\ff_1 \\
\ff_2 + i\ff_{M+2} \\
\ff_3+ i\ff_{M+3} \\
\vdots \\
\ff_{M+1} + i\ff_{2M+1} \\
\ff_{M+1} - i\ff_{2M+1} \\
\ff_{M} - i\ff_{2M} \\
\vdots \\
\ff_2 - i\ff_{M+2}
\end{pmatrix}
\end{align} 
where $\ff_j$ is the $j$th element of $\hat {\mathfrak f}$.  

To derive the transformed version of $\lam$ in the basis of $\hat {\mathfrak f}$, first decompose the matrix into blocks via
\begin{align}
\lam &= \begin{bmatrix}
A & B \\
C & D
\end{bmatrix}
\end{align}
where $A$ is $(M+1)\times (M+1)$, $B$ is $(M+1) \times M$, $C$ is $M \times (M+1)$, and $D$ is $M \times M$.  Clearly $A$ couples the first $M+1$ modes of $\f$ into one another, $B$ couples the remaining $M$ modes into the first $M+1$, and so forth.  However, because the signal remains real in the time-domain after transformation by $\lam$, knowledge of $A$ and $B$ is sufficient to determine the transformation matrix in the basis of $\hat {\mathfrak f}$.  Let $\FF$ denote that operator that reverse the order of columns in a matrix, $\mathcal T_{c}$ ($\mathcal T_{r}$) denote the operator that removes the first column (row) of a matrix, and $\boldsymbol 0_M$ denote a column vector with $M$ zeros.  Then it is straightforward to show that the operator $\lam$ transforms to 
\begin{align}
\boldsymbol \Theta \!&=\! \begin{bmatrix}
\tilde A & \tilde B \\
\tilde C & \tilde D
\end{bmatrix}
\end{align}
where
\begin{align}
\tilde A &= \Re(A) + \left[\boldsymbol 0_{M+1},\FF(\Re(B)) \right] \nonumber \\
\tilde B &= - \mathcal T_c(\Im(A)) + \FF(\Im(B)) \nonumber \\
\tilde C &= \mathcal T_r(\Im(A)) + \left[\boldsymbol 0_{M},\FF(\Im(B)) \right] \nonumber \\
\tilde D &= \mathcal T_r(\Re(A)) - \FF(\Re(B))
\end{align}
and $\Re$ and $\Im$ denote the real and imaginary components.  Thus we arrive at an expression for $\hat {\boldsymbol {\rm m}}$ expressed in the basis of $\hat {\mathfrak f}$, viz,
\begin{align}
\hat {\mathfrak m} = \sum_{j} \boldsymbol \Theta^{-1}_j [\bar {\mathfrak f} + \Delta \hat {\mathfrak f}_j]. \label{eq:goodbasis}
\end{align}

Minimizing Eq.\ \eqref{eq:appdoubobj} subject to Eq.\ \eqref{eq:goodbasis} may entail optimizing over $\mathcal O(100)$ or more variables corresponding to: (i) the scale parameters $c$, $v$, $R$, and $\Delta t$; and (ii) the real and imaginary parts of the mode-weights.  The latter comprise the majority of variables, although they only appear up to second order.  In contrast, the transformation variables, while few in number, appear in highly non-linear function associated with the matrix $\Th^{-1}$ in Eq.\ \eqref{eq:goodbasis}.  Further compounding these issues is the fact that both $\Th^{-1}$ and $\Xi$ are dense matrices, the latter possibly having eigenvalues close to zero.  This may yield a relatively large numerical problem that is poorly scaled, and thus challenging to solve.

Fortunately, the constraint given by Eq.\ \eqref{eq:appmultconstraint} is linear in the mode-weights, which yields a key simplification.  Without loss of generality, one finds
\begin{align}
\Delta \mathfrak f_1 = \Th_1\left[\hat {\mathfrak m} - \Th_1^{-1}\bar{\mathfrak f} - \sum_{j=2}^{\mathcal M}\Th_j^{-1}[\bar{\mathfrak f} + \Delta \hat {\mathfrak f}_j] \right]. \label{eq:applinear}
\end{align}
Equation \eqref{eq:applinear} can be substituted into Eq.\ \eqref{eq:doubobj} and minimization performed over the remaining modes $\Delta \hat {\mathfrak f}_j$ for $j\ge 2$ for fixed transformation parameters associated with the $\Th_j$.  We leave this exercise for the reader.  For the case of doublets, one find that 
\begin{align}
G&= \Th_1\left[\hat {\mathfrak m} - (\Th_1^{-1}+\Th_2^{-1})\bar{\mathfrak f} \right ] \nonumber \\
\Delta \hat {\mathfrak f}_2^* &= \Big[(\Th_1\Th_2^{-1})^{\rm T}\Xi^{-1}\Th_1\Th_2^{-1} + \Xi^{-1}  \Big]^{-1}\Th_1\Th_2^{-2}\Xi^{-1}G \nonumber \\
\Delta \hat {\mathfrak f}_1^* &= \Th_1[G - \Th_2^{-1}\Delta \hat {\mathfrak f}_2], \nonumber
\end{align}
where $\Delta \hat {\mathfrak f}_2^*$ and $\Delta \hat {\mathfrak f}_1^*$ are the optimal mode perturbations (the $*$ is distinct from the complex conjugate $\star$).  Having the $\Delta \hat {\mathfrak f}_j^*$ in terms of the transformation parameters (via the $\Th_j$), we may then express the objective as
\begin{align}
\mathcal L_d = \sum_j \left[\Delta \hat {\mathfrak f}_j^*\right]^{\rm T} \Xi^{-1} \Delta \hat {\mathfrak f}_j^* + \Delta \chi_j^{\rm \T}\Upsilon^{-1} \Delta \chi_j.
\end{align}
This $\mathcal L_d$ can then be optimized as a function of the scale transformations parameters.

\bibliography{metrics.bib}

\end{document}